\documentclass[twocolumn]{article}

\usepackage{amssymb}
\usepackage{amsmath}
\usepackage[hidelinks]{hyperref}
\usepackage{color}
\usepackage{pifont}
\usepackage{tikz}
\usepackage{mathrsfs}
\usepackage{graphicx}

\newtheorem{remark}{Remark}

\definecolor{bleuONERA}{RGB}{16,97,169}
\definecolor{grisONERA}{RGB}{64,64,66}
\usepackage{bbold}
\usepackage{xparse}
\usepackage{amsmath}
\usepackage{enumitem}
\usepackage{xcolor}

\usepackage{algorithm}
\usepackage{algorithmic}

\makeatletter \newcommand{\IfNoValueOrEmptyTF}[3] {\IfNoValueTF{#1}{#2}{\def\@tempa{#1}\ifx\@tempa\@empty#2\else#3\fi}} \makeatother

\DeclareDocumentCommand \rs {o} {%
  \IfNoValueTF{#1}{\mathbb{R}}{\mathbb{R}^{#1}}%
}
\DeclareDocumentCommand \cs {o} {%
  \IfNoValueTF{#1}{\mathbb{C}}{\mathbb{C}^{#1}}%
}

\newcommand{\udisk}{\mathcal{D}}
\newcommand{\ucir}{\partial \mathcal{D}}
\newcommand{\udiskcomp}{\overline{\mathcal{D}}}
\newcommand{\cp}{\mathbb{C}_+}

\newcommand{\chinf}{\mathcal{H}_\infty (\cp)}
\newcommand{\dhinf}{\mathcal{H}_\infty (\udiskcomp)}

\usepackage{tikz}
\usetikzlibrary{positioning,calc}
\usetikzlibrary{arrows,decorations.pathmorphing,backgrounds,fit,positioning,shapes.symbols,chains}
\usetikzlibrary{patterns,snakes}

\newcommand{\mat}[2]{\left [ \begin{array}{#1} #2 \end{array}\right ]}

\date{}

\begin{document}
\title{Discretisation of continuous-time linear dynamical model with the Loewner interpolation framework}

\author{
P. Vuillemin$^\dagger$, and
C. Poussot-Vassal$^\dagger$\\
\small $^\dagger$ONERA / DTIS, Universit\'e de Toulouse, F-31055 Toulouse, France
}


\maketitle

\begin{abstract}
An interpolation method for discretising continuous-time Linear Time Invariant (LTI) models is proposed in this paper. It consists first in using the Loewner interpolation framework on a specific set of frequency data and secondly to project the resulting model onto a stable subspace. The order of the discretised model may be chosen larger than the initial one thus allowing for trading complexity for accuracy if needed. Numerical examples highlight the efficiency of the method at preserving a satisfactory matching both in magnitude and phase in comparison to standard discretisation methods like ZOH or Tustin.
\end{abstract}


\section{Introduction}

In automatic control and dynamical systems theory, two different domains coexist: the continuous-time and discrete-time domains. While most of the tools available in one domain have a counterpart in the other, it is not unusual that a specific application requires to switch from one domain to the other.

\begin{figure}
    \centering
    \includegraphics[width=0.8\linewidth]{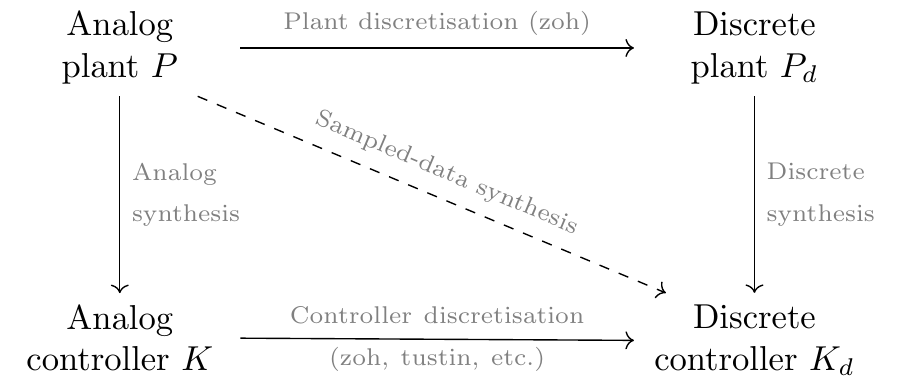}
    \caption{Different paths to design a discrete control-law $K_d$ from an analog plant $P$.}
    \label{fig:synth-scheme}
\end{figure}

In particular, for control-oriented applications, engineers often start from a physical model expressed in continuous-time to be used for the design of a control law. The latter is generally aimed at being implemented on a computer, which intrinsically evolves in discrete-time. As illustrated schematically in Figure \ref{fig:synth-scheme}, three approaches can be considered to address this issue:

\begin{enumerate}[label={\textit{(\roman*)}}]
    \item one may discretise the analog plant $P$ and then design a discrete control-law $K_d$,
    \item conversely, one may design an analog control-law $K$ and then discretise it,
    \item or, using dedicated techniques from sampled-data systems theory (see e.g. \cite[chap. 12-13]{chen1995optimal}), one may directly synthesise $K_d$ from $P$.
\end{enumerate}

While the direct nature of the latter method is appealing, it requires dedicated theoretical and numerical tools that are not as widespread as usual ones, especially in the industry where this approach would require their whole control design and analysis process to be rethought. For these reasons, one focuses here on the indirect methods that remain of practical interest.

For both indirect approaches, a discretisation step is required. This may have a detrimental impact with respect to the expected dynamical behaviour, especially if hard computational constraints on the sampling period are imposed by the technology, which may happen with critical systems. In that context, the availability of an efficient discretisation method is of particular interest and in this article, an approach based on the Loewner interpolatory framework \cite{Mayo:2007} is presented. It offers an interesting alternative to usual discretisation processes as it enables to reach a better frequency and time-domain matching with the continuous-time model.

\subsection{Problem statement \& contributions}
Let us consider a stable (see Remark \ref{rq:unstable}) continuous-time Linear Time Invariant (LTI) dynamical system described by the following state-space representation,
\begin{equation}
    \begin{array}{rcl}
         \dot x(t)&=& A x(t) + B u(t)  \\
         y(t)& = & C x(t) + D u(t)
    \end{array}
    \label{eq:sysc}
\end{equation}
where $x(t) \in \rs[n]$, $u(t) \in \rs[n_u]$ and $y(t) \in \rs[n_y]$ and its associated transfer function $G(s) = C(sI_n - A)^{-1} B + D$ which is assumed to lie in $\mathcal{L}_\infty (j\rs)$ and to be analytic in the open right half plane $\cp$, i.e. $G \in \chinf$.

The objective here is to determine, for a fixed time-step $h > 0$, a discrete-time model represented by the recurrence state-space equation,
\begin{equation}
    \begin{array}{rcl}
         x_d[k+1]&=& A_d x_d[k] + B_d u_d[k]  \\
         y_d[k]& = & C_d x_d[k] + D_d u_d[k]
    \end{array}
    \label{eq:sysd}
\end{equation}
where $x_d[k] \in \rs[n_d]$, $u_d[k] \in \rs[n_u]$, $y_d[k] \in \rs[n_y]$ and such that:
\begin{enumerate}[label={\textit{(o\arabic*)}}]
    \item its associated transfer function $G_d(z) = C_d(zI_{n_d} - A_d)^{-1} B_d + D_d$ is stable, i.e. $G_d \in \dhinf$,
    \item the input-ouput behaviour of $G$ is well reproduced by $G_d$.
\end{enumerate}

Existing discretization methods such as Tustin are clearly able to build a discrete-time model satisfying \textit{(o1)} but the input-output behaviour may be quite far from the original one when $h$ is too large. In this article, a method is proposed to build such a model by first using the Loewner framework to interpolate a specific set of frequency data and then projecting the resulting model onto the stable subspace $\dhinf$ to enforce stability.

\begin{remark}
  \label{rq:unstable}
 The unstable case may be of interest, e.g. for the discretisation of a control-law with an integrator. In that case, the stable and anti-stable parts of the model should be splitted and the process described in this paper should be applied only to discretise the stable part. The unstable part may be discretised independently, e.g. with standard methods or by substituting the final projection subspace by $\mathcal{H}_\infty(\udisk)$ in the proposed approach. Note however that the time-domain discretisation error may be unbounded for unstable systems.
\end{remark}

In section \ref{sec:measure}, elements from sampled-data systems theory are recalled to highlight the difficulty underlying the objective \textit{(o2)} and how one may actually quantify the discretisation error. Then in section \ref{sec:loewn}, the proposed discretisation method is detailed. Two numerical applications that highlight the performances of the approach are then presented in section \ref{sec:num}. Finally, section \ref{sec:ccl} concludes and draw some further possible improvments.

\subsection{Notations}
Let us denote by $\rs$ the set of real numbers, $\mathbb{C}$ the set of complex numbers, $\cp$ the open right half plane, $\udisk$ the open unit disk, $\ucir$ its boundary and $\udiskcomp$ the complementary of the closed unit disk, respectively.

Let $\mathcal{L}_2(\mathcal{I})$ ($\mathcal{I} = \mathbb{R},\, \ucir$) be the set of functions that are square integrable on $\mathcal{I}$. Let $\mathcal{H}_2(\mathcal{\udisk})$ (resp. $\mathcal{H}_2(\mathcal{\udiskcomp})$) be the subset of $\mathcal{L}_2(\ucir)$ containing the functions analytic in $\udisk$ (resp. $\udiskcomp$). Similarly, let $\mathcal{L}_\infty (\mathcal{I})$ ($\mathcal{I} = j\mathbb{R}, \, \ucir$) be the set of functions that are bounded on $\mathcal{I}$. Let $\mathcal{H}_\infty(\mathcal{\udisk})$ (resp. $\mathcal{H}_\infty(\mathcal{\udiskcomp})$) be the subset of $\mathcal{L}_\infty(\ucir)$ containing the functions analytic in $\udisk$ (resp. $\udiskcomp$) and $\mathcal{H}_\infty(\cp)$ the subset of $\mathcal{L}_\infty(j\rs)$ of functions analytic in $\cp$.
The Fourier transform of a time-domain signal $v(t) \in \mathcal{L}_2(\mathbb{R})$ is denoted by $\hat{v} = \mathcal{F}(v)$.


\section{Measure of the discretisation error}
\label{sec:measure}
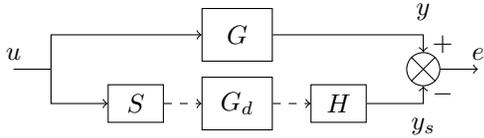
\begin{figure}
\centering
\begin{tikzpicture}
  \node (G) [anchor = west, draw, minimum height= 0.7cm, text width = 0.7cm, text centered ]{$G$};
  \node (gd) at ($(G.south)+(0,-0.2cm)$) [anchor = north, draw,minimum height= 0.7cm, text width = 0.7cm, text centered ] {$G_d$};
  \node (st) at ($(gd.west) + (-0.5cm,0)$) [anchor = east, draw,minimum height= 0.5cm, text width = 0.5cm, text centered ] {$S$};
  \node (ht) at ($(gd.east) + (0.5cm,0)$) [anchor = west, draw,minimum height= 0.5cm, text width = 0.5cm, text centered ] {$H$};

  \node (add) [draw, circle, text width=4pt] at ($0.5*(G.east) + 0.5*(gd.east) + (2cm,0)$) {};

 \node (start) at ($0.5*(G.west) + 0.5*(gd.west) - (2cm,0)$) {};
  \draw [->] (start.center) |- (G.west);
  \draw [->] (start.center) |- (st.west);
  \draw [->] (G.east) -| (add.north) ;
 \draw [->] (ht.east) -| (add.south) ;
  \draw [dashed, ->] (st.east) -- (gd.west);
  \draw [dashed, ->] (gd.east) -- (ht.west);
  \draw [-] (start.center) --+ (-0.5cm,0) node [above] {$u$};
  \draw [->] (add.east) --+ (0.5cm,0) node [above] {$e$};
  \node [anchor=west] at ($(add.south)+(0,-0.1cm)$) {$-$};
        \node [anchor=west] at ($(add.north)+(0,0.1cm)$) {$+$};

  \node [anchor=north] at ($(add.south)+(0,-0.3cm)$) {$y_s$};
  \node [anchor=south] at ($(add.north)+(0,0.3cm)$) {$y$};

  \draw (add.45)--(add.225);
  \draw (add.135) -- (add.315);
\end{tikzpicture}
\caption{Interconnection for the measurement of the discretisation error between $G$ and $G_d$.}
\label{fig:discretisation_error}
\end{figure}

Quantifying the error induced by the discretisation of $G$ is not trivial due to the incompatible nature with $G_d$ that prevents from interconnecting the two systems. Indeed, one cannot just consider $G-G_d$ and digital to analog converters are required. For a sampling time $h>0$, the latter are modelled here as the ideal sampler $S$ and holder $H$. Then, the error system is given by the interconnection of Figure \ref{fig:discretisation_error} where discrete-time signals are represented by dashed lines. Such an interconnection of continuous and discrete time models is called a Sampled-Data System (SD).

The model $\tilde{G} = S G_d H$ is not LTI but $h$-periodic and consequently, so is the error $G - \tilde{G}$. Usual system norms can therefore not directly be applied  to quantify $G - \tilde{G}$.  This problem has been addressed in the literature for direct SD synthesis (see e.g.\cite{bamieh:lifting:1991,chen1995optimal} and references therein). The recurring idea is to use continuous lifting to transform a periodic system into a discrete-time LTI model with infinite input and output spaces on which equivalent $\mathcal{H}_2$ or $\mathcal{H}_\infty$ norms can be defined.

While this framework appears well suited to evaluate $G - \tilde{G}$, the infinite dimensionality of the lifted model makes it quite technical. That is why here, a more straightforward approach based on the frequency characterisation of the discretisation error formulated in \cite[sec.3.5]{chen1995optimal} is considered instead. For that purpose, models for the ideal sampler and holder are recalled in section \ref{ssec:idealsh} and the frequency error is presented in section \ref{ssec:freqerr}.

Note that in \cite{chen1995optimal}, the authors use the $\lambda$-transform while here, one considers the $z$-transform, the sign of the exponential in the Fourier transforms of discrete signal is thus modified.

\subsection{Ideal sampler and holder}
\label{ssec:idealsh}

Let us consider a continuous-time signal $v(t)$ and the sampling period $h$. The ideal sampler $S$ transforms $v(t)$ into a discrete sequence $v_d[k]$ such that,
\begin{equation}
    v_d[k] = v(kh),\, k\in\mathbb{Z}.
\end{equation}
As shown in Lemma 3.3.1 of \cite{chen1995optimal}, the Fourier transforms $\hat{v}_d =\mathcal{F}(v_d)$ and $\hat{v} = \mathcal{F}(v)$ are linked as follows,
\begin{equation}
    \hat{v}_d(e^{j\omega h}) = \frac{1}{h} \sum_{k\in\mathbb{Z}} \hat{v}(j \omega + j k \omega_s)
    \label{eq:freq_sampler}
\end{equation}
where $\omega_s = 2\pi/h$ is the sampling frequency. Equation \eqref{eq:freq_sampler} hightlights the frequency aliasing phenomena since all the multiples of the sampling frequency are indistinguishable in the output. Note that the sampling operator $S$ is not bounded for any signal in $\mathcal{L}_2(\mathbb{R})$ as shown in \cite{chen1991input}. To be bounded, the input signal $v$ must be restricted to the class of bandlimited $\mathcal{L}_2 (\mathbb{R})$ signals or it must be filtered by a finite-dimensional stable and strictly causal system.

Similarly, the holder $H$ transforms a sequence $v_d[k]$ into a continuous-time signal $v(t)$ such that,
\begin{equation}
    v(t) = v_d[k],\quad kh\leq t < (k+1)h.
\end{equation}
The impulse response of the holder can be defined as the difference of two unit steps delayed by $h$. Let $R(s) = \frac{1 - e^{-sh}}{sh}$ be the associated transfer function. Then, as shown in Lemma 3.3.2 of \cite{chen1995optimal}, the Fourier transforms of $v$ and $v_d$ are linked as follows,
\begin{equation}
    \hat{v}(j \omega) = h R(j\omega) \hat{v}_d(e^{j\omega h}).
    \label{eq:freq_holder}
\end{equation}
With respect to Figure \ref{fig:discretisation_error}, coupling equations \eqref{eq:freq_sampler} and \eqref{eq:freq_holder} enables to express the frequency-domain relationship between $u$ and $y_s$,
\begin{equation}
    \hat{y}_s(j\omega) =  R(j\omega) G_d(e^{j\omega h}) \sum_{k\in \mathbb{Z}} \hat{u}(j\omega + jk \omega_s).
    \label{eq:freq_relation}
\end{equation}

\subsection{A frequency domain error}
\label{ssec:freqerr}

Using equation \eqref{eq:freq_relation}, one can express the frequency-domain relationship between $u$ and the discretisation error $e = y - y_s$ from Figure \ref{fig:discretisation_error} as
\begin{equation}
    \hat{e}(j\omega) = G(j\omega) \hat{u}(j\omega) - R(j\omega) G_d(e^{j\omega h})\sum_{k\in\mathbb{Z}} \hat{u}(j\omega + jk\omega_s).
    \label{eq:freq_err}
\end{equation}
Assuming that $u$ is bandlimited, i.e. that $\hat{u}(j\omega) = 0$ for $|\omega| > \omega_s/2 = \omega_N$, then \eqref{eq:freq_err} becomes
\begin{equation}
    \hat{e}(j\omega) = \left ( G(j\omega) - R(j\omega) G_d(e^{j \omega h}) \right ) \hat{u}(j\omega) .
    \label{eq:simp_freq_err}
\end{equation}
This readily suggests to consider
\begin{equation}
    e_{\infty}(G,G_d) =  \max_{\omega < \omega_N} \left | G(j\omega) - R(j\omega) G_d(e^{j\omega h}) \right |,
    \label{eq:err}
\end{equation}
to quantify the discretisation error. This frequency-domain characterisation of the error inspired the discretisation process presented in the next section. For the Multiple Input Multiple Output (MIMO) case, the absolute value in \eqref{eq:err} may be replaced by the $2$-norm.

\section{Loewner interpolation for discretisation}
\label{sec:loewn}
The proposed strategy relies on a specific use of the Loewner interpolation framework. The latter is recalled in section \ref{ssec:loe}. The main idea for Loewner-driven discretisation is then given in section \ref{ssec:dloe} and stability issues of the discretised model are discussed in section \ref{ssec:stab}. Finally, section \ref{ssec:alg} sums-up the proposed approach.

\subsection{Reminder of the Loewner framework}
\label{ssec:loe}

The main elements of the Loewner framework are recalled thereafter in the single-input single-output (SISO) case and readers may refer to \cite{Mayo:2007} for a complete description and extension to the MIMO one.

The Loewner approach is a data-driven method aimed at building a rational descriptor LTI dynamical model $G_d^m$ of dimension $m$  which interpolates given frequency-domain data. More specifically, let us consider a transfer function $G$ and a set of distinct interpolation points  $\{ z_i \}_{i=1}^{2m} \subset \cs$ which is split in two equal subsets as
\begin{equation}
\{z_i\}_{i=1}^{2m} =\{\mu_i\}_{i=1}^{m} \cup \{\lambda_i\}_{i=1}^{m}.
\end{equation}
The method then consists in building the Loewner and shifted Loewner matrices defined as,
\begin{equation}
\scriptsize
    \left [ \mathbb{L} \right ]_{ij} = \frac{G(\mu_i) - G(\lambda_j)}{\mu_i - \lambda_j}
    \text{ and }
        \left [ \mathbb{L}_s \right ]_{ij} = \frac{\mu_i G(\mu_i) - \lambda_j G(\lambda_j)}{\mu_i - \lambda_j}
\end{equation}
The model $G_d^m$ that interpolates $G$ is then given by the following descriptor realisation,
\begin{equation}
\Sigma_d^m:\left \lbrace
\begin{array}{rcl}
E_d  x_{k+1} &=& A_d x_k + B_d u_k\\
y_k &=&C_d x_k
\end{array}
\right .
\label{eq:descr}
\end{equation}
where $E_d = -\mathbb{L}$, $A = -\mathbb{L}_s$, $[B_d]_i = G(\mu_i)$ and $[C_d]_i = G(\lambda_i)$ ($i=1,\ldots,m)$.

Assuming that the number $2m$ of available data is large enough, then  it has been shown in \cite{Mayo:2007} that a minimal model $G_d^r$ of dimension $r < m$ that still interpolates the data can be built with a projection of \eqref{eq:descr} provided that, for $i=1,\ldots,2m$,
\begin{equation}
    rank(z_i \mathbb{L} - \mathbb{L}_s) = rank( [\mathbb{L}\,\mathbb{L}_s]) = rank([\mathbb{L}^T\, \mathbb{L}_s^T]^T) = r.
    \label{eq:rankCond}
\end{equation}
In that case, let us denote by $Y \in \cs[m \times r]$ the matrix containing the first $r$ left singular vectors of $[\mathbb{L}\, \mathbb{L}_s]$ and $X \in \cs[m \times r]$  the matrix containing the first $r$ right singular vectors of $[\mathbb{L}^T\, \mathbb{L}_s^T]^T$. Then,
\begin{equation}
 \scriptsize
    E_d^r = Y^H E_d X,\,
    A_d^r = Y^H A_d X,\,
    B_d^r = Y^H B_d,\,
    C_d^r = C_d X,
    \label{eq:proj}
\end{equation}
is a realisation of this model $G_d^r$ with a McMillan degree equal to $rank(\mathbb{L})$.

Note that if $r$ in \eqref{eq:rankCond} is superior to $rank(\mathbb{L})$ then $G_d^r$ can either have a direct-feedthrough $D_d\neq 0$ or a polynomial part. In the reminder of the paper, one assumes that the latter case does not happen so that $G_d^r$ can be described by a state-space realisation as in \eqref{eq:sysd}.

The number $k$ of singular vectors used to project the system $G_d^m$ may be decreased even further than $r$ at the cost of approximate interpolation of the data. This allows for a trade-off between complexity of the resulting model and accuracy of the interpolation.

\subsection{Application of the Loewner framework for discretisation}
\label{ssec:dloe}

Let us consider the transfer function $G \in \chinf$ associated with the LTI model described by \eqref{eq:sysc} to be discretised at the sampling time $h$.

To build a $k$-th order discrete-time model $G_d^k$ that matches the input-output behaviour of $G$, the frequency-domain characterisation of the discretisation error \eqref{eq:err} suggests that $G_d^k$ must be such that
\begin{equation}
    R(j\omega) G_d^k(e^{j\omega h}) = G(j\omega),
    \label{eq:interp_cond}
\end{equation}
for $|\omega| < \omega_N$. Equation \eqref{eq:interp_cond} represents an infinite number of interpolation conditions that may be approximated by sampling the interval $[0,\omega_N]$ such that for $i=1,\ldots,2m$,
\begin{equation}
    R(j\omega_i) G_d^k(e^{j\omega_i h}) = G(j\omega_i).
    \label{eq:sampled_interp_cond}
\end{equation}
Such a model can be built by applying the Loewner interpolation framework recalled in section \ref{ssec:loe} to the following set of frequency data,
\begin{equation}
    \left \lbrace e^{j\omega_i h}, R(j\omega_i)^{-1} G(j\omega_i) \right \rbrace_{i=1}^{2m}.
    \label{eq:dataset}
\end{equation}
Obviously, should the order $k$ be lower than the McMillan order $r$ of the exact interpolating model given by the Loewner approach, then the interpolation conditions \eqref{eq:sampled_interp_cond} will not be perfectly satisfied.




\subsection{About the stability of the discretised model}
\label{ssec:stab}

The Loewner framework does not ensure the stability of the resulting interpolant model. Therefore, even if $G \in \chinf$, the transfer  function $G_d^k$ obtained via the process described  above may not lie in $\dhinf$ which is a major drawback in comparison to stability preserving discretisation schemes like the Tustin one.

To overcome this issue, one can apply the same process as in \cite{gosea:stability:2016}. It consists in projecting the unstable model $G_d^k$ onto $\dhinf$ so that the $\mathcal{L}_\infty$ norm between $G_d^k$ and its projection is minimised. This is the so-called Nehari problem for which solutions have been given in the continuous-time domain \cite{Glover:1984} and in the discrete-time domain, see e.g. \cite{mari:modifications:2000} .

Let us denote by $P_\infty$ this projection operator so that if $G_d^k \in \mathcal{L}_\infty(\ucir) $ then $P_\infty(G_d^k) \in \dhinf$ minimises $\Vert G_d^k - P_\infty(G_d^k) \Vert_{\mathcal{L}_\infty}$. Note that the order of $P_\infty(G_d^k)$ depends on the number of unstable poles of $G_d^k$ and is lower than $k$ when $G_d^k$ is unstable. In particular, if $G_d^k$ has $k_s$ and $k_u$ stable and unstable poles and that $q$ is the multiplicity of the largest unstable Hankel singular value, then $P_\infty(G_d^k)$ has order $k_s + k_u - q$ (see \cite{mari:modifications:2000}). To avoid this issue and for a numerically more robust approach, sub-optimal projection methods may alternatively be considered (see e.g. \cite{Kohler:2014}).



\begin{remark}
\label{rq:ldproj}
A similar problem can be considered using the $\mathcal{L}_2$-norm and is actually much easier to solve considering the decomposition $\mathcal{L}_2(\ucir) = \mathcal{H}_2(\udisk) \oplus \mathcal{H}_2(\udiskcomp)$. The solution is simply obtained by discarding the unstable part of the model. However this solution has generally a greater impact on the frequency behaviour of the model which is not desirable here.
\end{remark}

\subsection{Loewner-driven discretisation}
\label{ssec:alg}

\begin{algorithm}[t]
\caption{Loewner-driven discretisation} \label{theAlg}
\begin{algorithmic}[1]
\REQUIRE A continuous-time model $G \in \chinf$, a sampling time $h>0$, an upper bound $\bar{k}$ of the desired order and a number $m$ of interpolation points.
\STATE Sample the interval $]0,\omega_N[$ in $2m$ points $\omega_i$
\STATE Evaluate $R(j\omega_i)^{-1} G(j\omega_i)$
\STATE Apply Loewner to the data-set \eqref{eq:dataset} to get a first model $G_d^m$ that matches all the data and get the underlying minimal order $r$
\STATE Set $k=min(r, \bar{k})$
\STATE Reduce $G_d^m$ to $G_d^k$ as in \eqref{eq:proj}
\STATE Project $G_d^k$ onto a stable subspace, i.e. compute $G_d = P_\infty(G_d^k)$
\RETURN the model $G_d\in\dhinf$ of order $\leq \bar{k}$
\end{algorithmic}
\end{algorithm}

The process for Loewner-driven discretisation is summarised in Algorithm \ref{theAlg}. The following comments can be added:
\begin{itemize}
    \item the reduction step from $G_d^m$ to $G_d^k$ (step $5$) may also be achieved by standard model approximation techniques \cite{AntoulasBook:2005}. In that case, it would be preferable to first project $G_d^m$ to $G_d^r$ as in equation \eqref{eq:proj} which does not induce any detrimental effect on the interpolation. Then $G_d^r$ should be projected on $\dhinf$ and finally a stability preserving model reduction method could be used to obtain $G_d^k$.
    \item As both $G_d^r$ and $G_d$ are easily available during the process, $\Vert G_d^r - G_d \Vert_{\mathcal{L}_\infty}$ can be evaluated to get an estimation of the discretisation error $e_\infty(G,G_d)$ in \eqref{eq:err}. \item Consequently, when $k = \bar{k} < r$, computing $\Vert G_d^r - G_d^k \Vert_{\mathcal{L}_\infty}$ gives insight on whether the maximal allowed order $\bar{k}$ is enough to ensure a low error.
    \item If $G_d^k$ is unstable, then step $6$ leads to a loss of order (see section \ref{ssec:stab}). Therefore, when $\bar{k} < r$ (which is likely to happen\footnote{Indeed, one tries to approximate data coming from an infinite dimensional model by a rational one. A large order $r$ is generally required to achieve an exact interpolation.}), $k$ should be increased above $\bar{k}$ so that $P_\infty(G_d^k)$ is of order $\bar{k}$ thus avoiding any loss of accuracy.
\end{itemize}

It should also be noted that Algorithm \ref{theAlg} does not exploit the state-space structure of $G$ and only needs frequency data from it. Therefore, the approach may be applied to a wider class of models than just those described by a state-space realisation as \eqref{eq:sysc}. This is illustrated in section \ref{ssec:tds} on a time-delay system.

\section{Numerical applications}
\label{sec:num}
The discretisation process described in Section \ref{sec:loewn} is analysed and compared to usual discretisation methods on a state-space model such as \eqref{eq:sysc} in section \ref{ssec:delayfree}. Its versatility is then illustrated in section \ref{ssec:tds} through the discretisation of a model with internal delays.

Some general remarks concerning the comparison process:
\begin{itemize}
\item The model is compared to the well-known methods Tustin, considered without prewarping, and ZOH (see e.g. \cite{aastrom:2013:computer}) and also to the impulse-invariant method \cite{kowalczuk:discrete:1993}. More specifically, their implementation in the routine \texttt{c2d} from the MATLAB's Control Toolbox is considered.
\item The frequency error \eqref{eq:err} is approximated by finely sampling the interval $\Omega = [10^{-3},\omega_N-10^{-3}]$ with $5\,000$ linearly spaced points. To ease the interpretation, the error is divided by the $\mathcal{H}_\infty$ norm of the continuous-time model to obtain a relative error, i.e.
\begin{equation}
\tilde{e}_\infty(G,G_d) =  e_\infty(G,G_d)/\Vert G \Vert_{\mathcal{H}_\infty}.
    \label{eq:relerr}
\end{equation}
\end{itemize}

\subsection{Discretisation of a delay-free LTI model}
\label{ssec:delayfree}

Let us consider the following $4$-th order continuous-time stable linear model
\begin{equation}
    G(s) = \frac{1 + 0.05 s/\sqrt{2} + s^2/2}{(1 + 0.1 s + s^2)(1+0.05 s/\sqrt{5} + s^2/5)}.
    \label{eq:ex1}
\end{equation}

\begin{figure}
    \centering
    \includegraphics[width=0.95\linewidth]{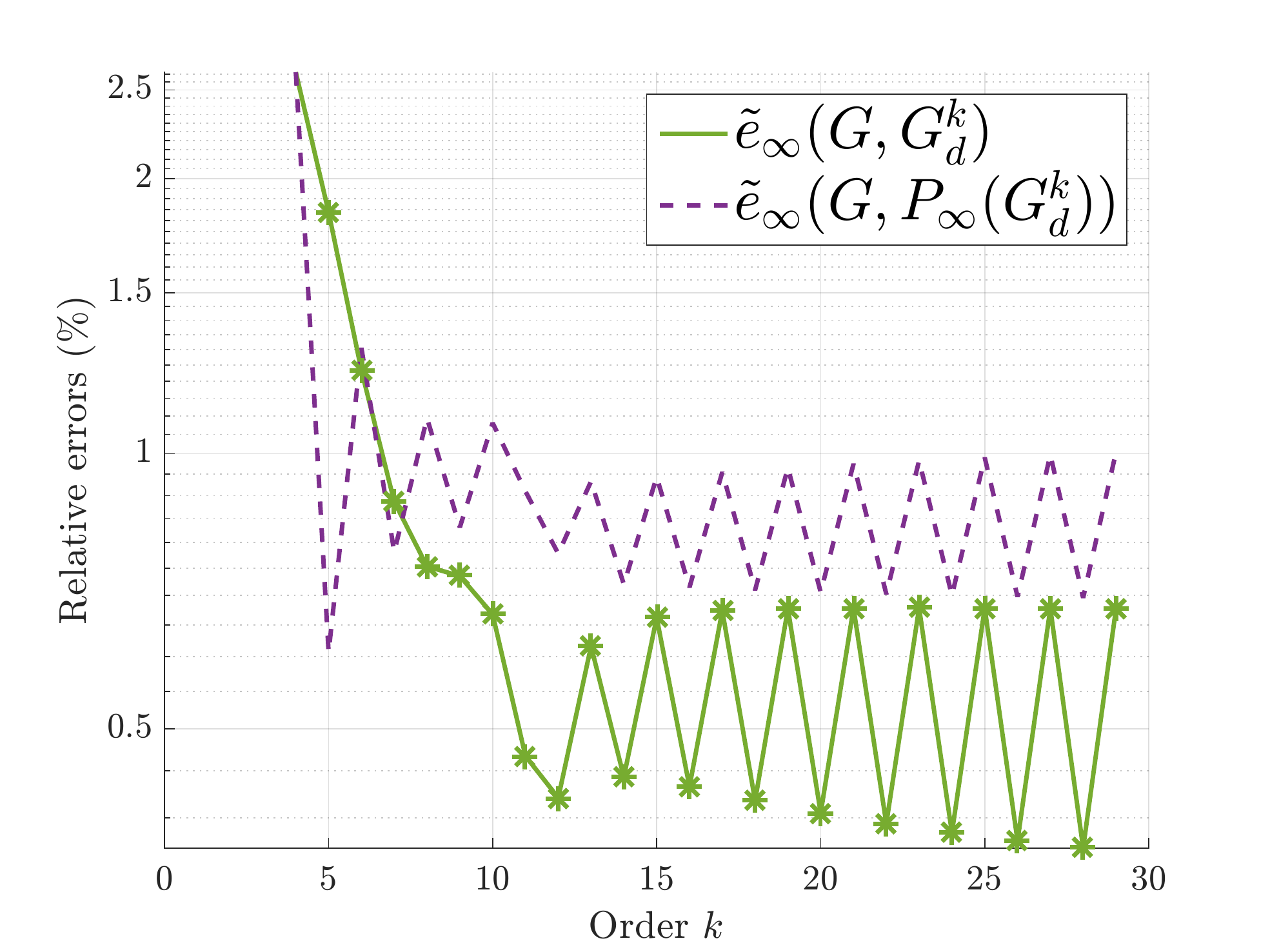}
    \caption{Relative frequency errors $\tilde{e}_\infty (G, G_d^k)$, $\tilde{e}_\infty(G, P_\infty(G_d^k))$ versus the order $k$. A star marker indicates an unstable model.}
    \label{fig:errors1}
\end{figure}

\emph{Analysis of the method. }The system \eqref{eq:ex1} is discretised to $G_d^k$ for a sampling time $h = 0.4s$ for increasing orders $k=4,\ldots,r$ with $100$ points linearly spaced in $\Omega$. The McMillan order $r$ of the minimal exact interpolation returned by the Loewner approach is $r=29$ here and $G_d^k$ is unstable as soon as $k>4$. The frequency errors \eqref{eq:relerr} are computed for $G_d^k$ and $P_\infty(G_d^k)$ to evaluate the impact of $k$ and of the stable projection step on the performances. The results are plotted in Figure \ref{fig:errors1}.  Note that the order of $P_\infty(G_d^k)$ is lower than $k$, by $1$ here.

First, one can notice that all the relative errors are quite low, below $3\%$. Increasing the order $k$ can indeed lead to better performances, however, the (slow) decrease is not monotonic and oscillations appear. By looking at $\tilde{e}_\infty(G,G_d^k)$, it seems that the curves goes up when $k$ is odd. Similar behaviour can be observed in model approximation when trying to reduce a model having an even number of poles with a model having an odd number of poles. This is supported by the fact that adding a real pole to $G(s)$ inverts the oscillation here. It is not clear why the phenomenon  only appears for $k>12$ though.

As expected, the projection onto the stable subspace tends to decrease the performances, excepted for some points where $\tilde{e}_\infty(G, P_\infty(G_d^k)) < \tilde{e}_\infty (G, G_d^k)$. Still, the two errors remain close. The best stable discretised model thus appear to be $P_\infty(G_d^5)$ which is of order $4$.

The error curves in Figure \ref{fig:discretisation_error} suggest that Algorithm \ref{theAlg} could be improved by looping over the order $k$ until $\bar{k}$ and retaining the projected model that minimises the interpolation error.

\emph{Comparison with other discretisation methods. }The stable models $G_d^4$ and $P_\infty(G_d^5)$, denoted $G_d$ in the sequel, have the following transfer functions,
\begin{equation}
\begin{array}{lcl}
G_d^4(z) &= &\frac{
    0.46194 (z-0.3987) (z^2 - 1.654z + 0.9954)}{
  (z^2 - 1.806z + 0.9606) (z^2 - 1.225z + 0.9562)}\\
    G_d(z)&= &\frac{
  0.17617 (z+1.347) (z-0.09051) (z^2 - 1.669z + 0.9737)}{
     (z^2 - 1.806z + 0.9607) (z^2 - 1.225z + 0.9562)}
    \end{array}
\end{equation}
Their associated errors are reported in Table \ref{table:errors} together with the errors obtained with other discretisation approaches. The invariant impulse approach, leading to $H_d^{ii}$, is clearly the most efficient method among the classic ones. However it still leads to an error magnitude larger than the one obtained with $G_d^4$ and $G_d$.


\begin{figure}
    \centering
    \includegraphics[width=0.95\linewidth]{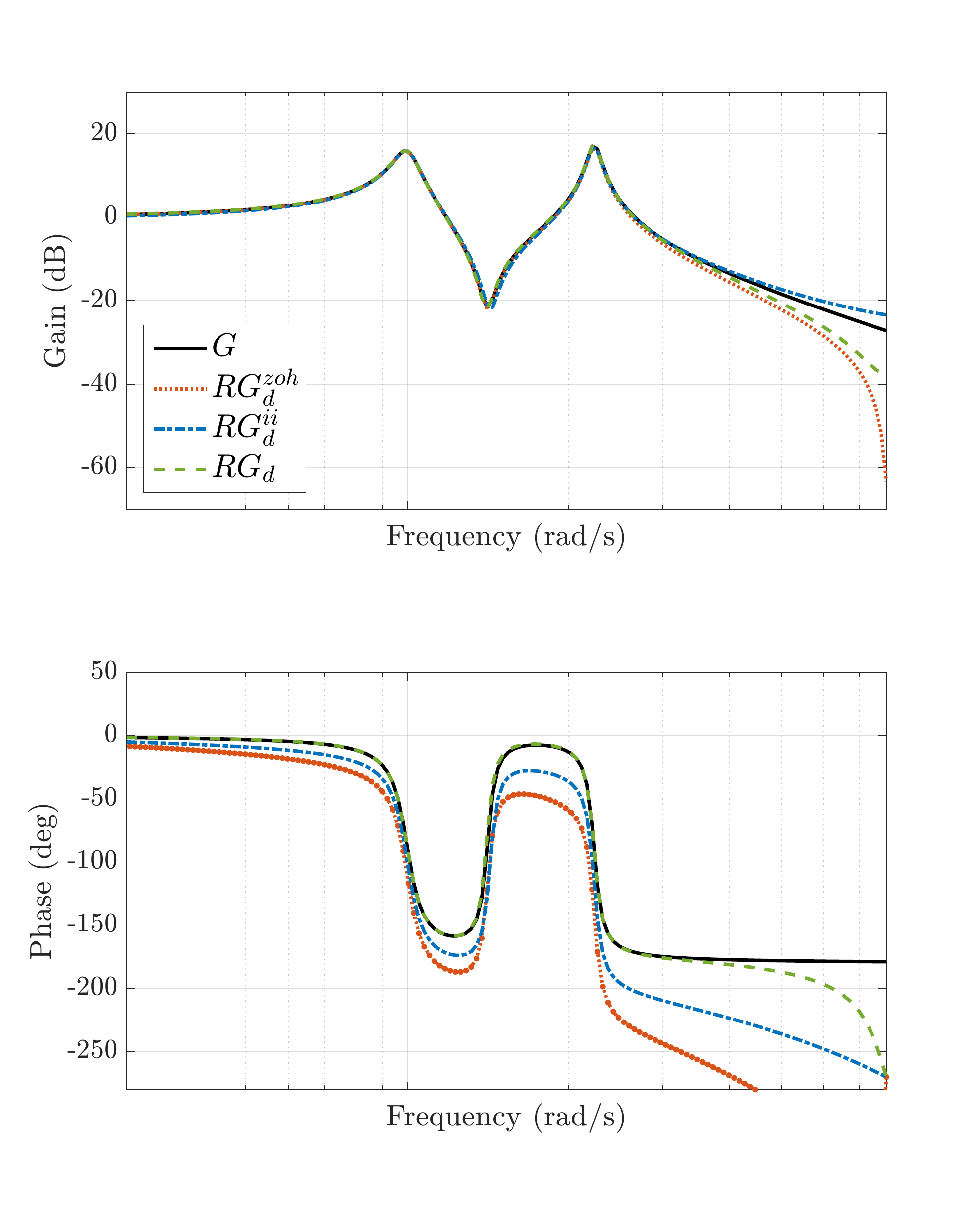}
    \caption{Frequency-domain responses (up to the Nyquist frequency) of the continuous-time model $G$ and its discretised counterparts $G_d^{zoh}$,  $G_d^{ii}$ and $G_d$.}
    \label{fig:bode}
\end{figure}

\begin{table}
\centering
\caption{Discretisation errors between the continuous-time model $G$ and its discretised counterparts. All models are of order $4$.}
\label{table:errors}
\scriptsize
\begin{tabular}{|c|ccccc|}
  \hline
  model  & $G_d^{tus}$ & $G_d^{zoh}$ & $G_d^{ii}$  & $G_d^4$ & $G_d$\\
  $	\tilde{e}_{\infty}(G,\cdot) (\%)$ & $113.46$& $83.88$&$44.19$&$2.61$& $\mathbf{0.61}$
  \\\hline
\end{tabular}
\end{table}

More visually, the frequency responses of $G$, $R G_d^{zoh}$, $R G_d^{ii}$ and $R G_d$ are plotted in Figure \ref{fig:bode}. All the discretised models are able to catch pretty accurately the magnitude of $G$, however, the phase is quite modified with ZOH and the impulse invariant methods while the proposed approach remains accurate up to $4$rad/s.

\begin{figure}
    \centering
    \includegraphics[width=0.95\linewidth]{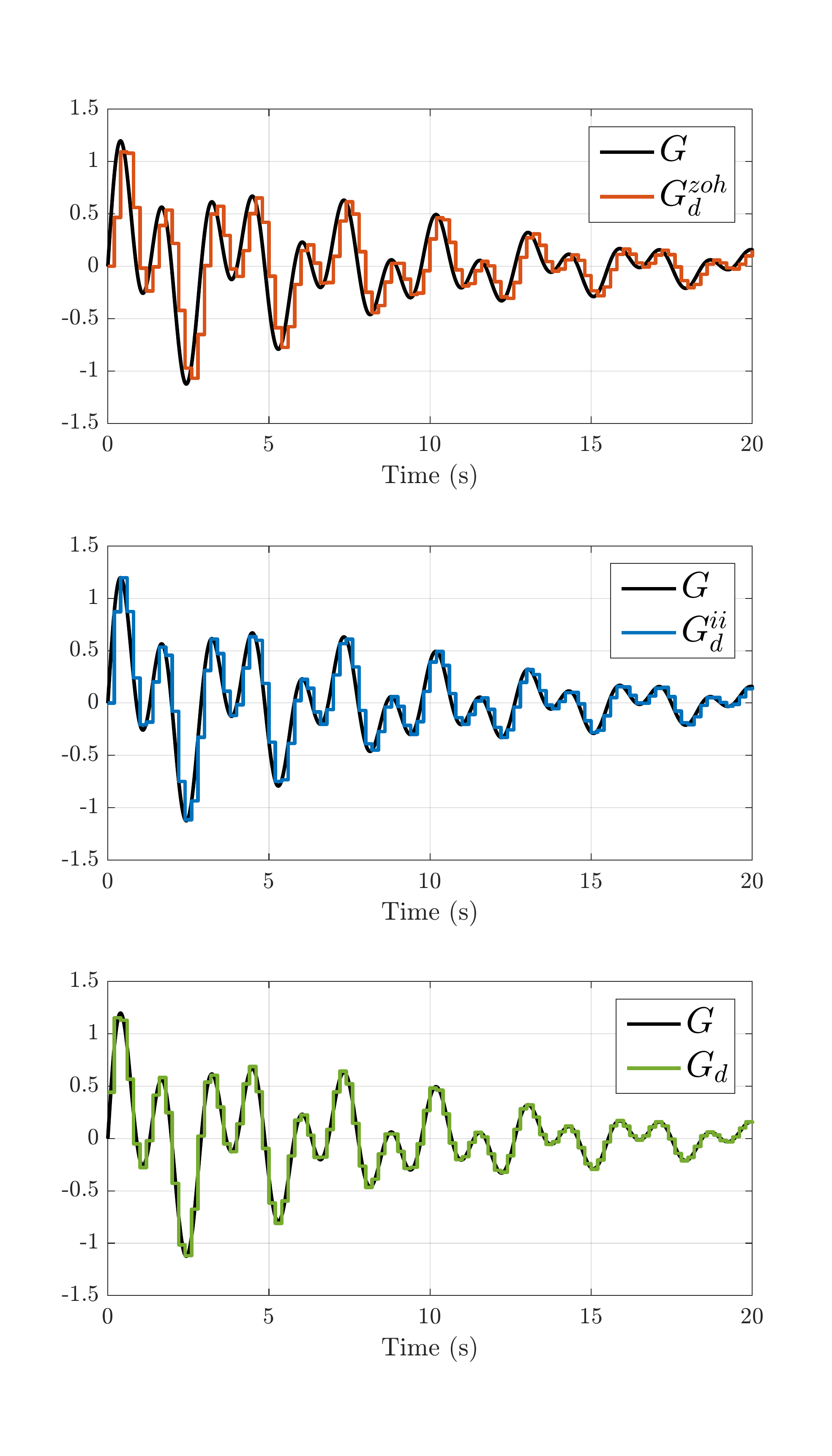}
    \caption{Impulse responses of $G$ and its discretised counterparts $G_d^{zoh}$ (top), $G_d^{ii}$ (middle) and $G_d$ (bottom).}
    \label{fig:ir}
\end{figure}

\begin{figure}
    \centering
    \includegraphics[width=0.95\linewidth]{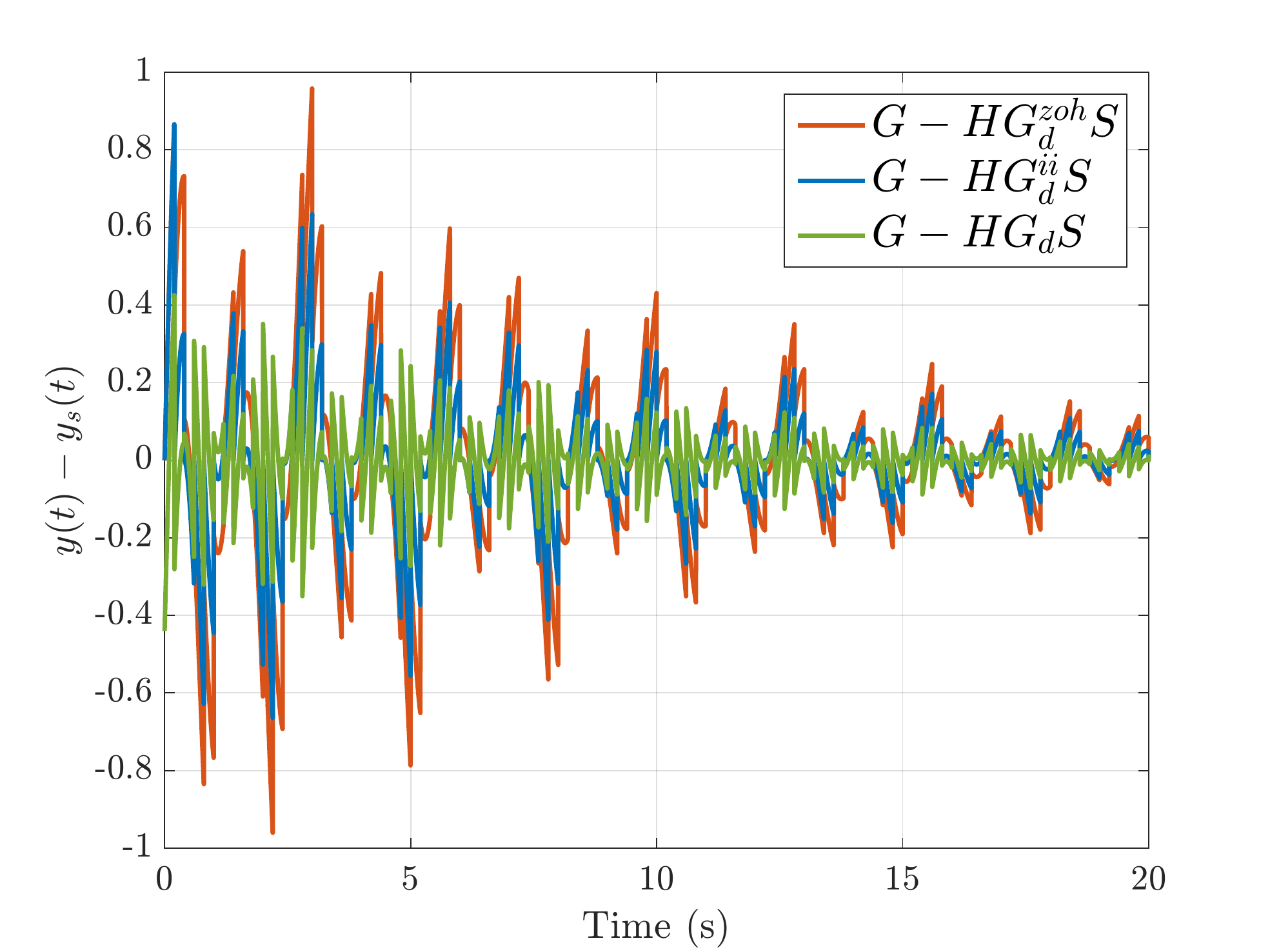}
    \caption{Time-domain error $y(t) - y_s(t)$ associated with the impulse responses of $G_d^{zoh}$, $G_d^{ii}$ and $G_d$.}
    \label{fig:tderror}
\end{figure}

To check the corresponding behaviour in time-domain, the impulse responses of the different models $G$, $G_d^{zoh}$, $G_d^{ii}$ and $G_d$ are compared. The resulting output signals are plotted in Figure \ref{fig:ir}. It highlights the phase distortion induced by the ZOH method. By design\footnote{Note that the impulse invariant method relies on the partial fraction decomposition of the transfer function and may therefore have trouble dealing with systems containing Jordan chains.}, the impulse invariant method matches the impulse response of the continuous-model at the sampling instants. On the contrary, the model $G_d$ appears to match the continuous response in-between sampling instants. This behaviour translates to a lower average time-domain error as shown in Figure \ref{fig:tderror}. This is confirmed when computing the relative $l_2$ errors of the sequence,

\begin{equation}
  e_2 = \Vert y-y_d\Vert_2 / \Vert y \Vert_2,
\end{equation}
which is equal to $72$\% for $G_d^{zoh}$, $42$\% for $G_d^{ii}$ and to $22$\% for $G_d$.

\subsection{Application to approximate discretisation of time-delay systems}
\label{ssec:tds}

As the Loewner discretisation process is based solely on frequency-domain data, it can actually be applied to a wider class of models than those described by state-space as \eqref{eq:sysc}. For instance, it can be used to get an approximate discretisation of time-delay systems (TDS).

As an illustration, let us consider the following TDS \cite{izmailov:analysis:1996} modelling a high speed network,
\begin{equation}
    \left \lbrace
    \begin{array}{l}
    \dot{x}(t) = A_0 x(t) + A_1 x(t - \tau) + A_2 x (t - \tau - \gamma) + B u(t)\\
    y(t) = C x(t)
    \end{array}
    \right .
    \label{eq:tds}
\end{equation}
where
\begin{equation}
\tiny
A_0 = \mat{cc}{0&0\\1&0},\, A_1 = -2A_0^T,\, A_2 = -1.75 A_0^T,\, B =\mat{c}{1\\0},\,C = \mat{c}{0\\1}^T
\end{equation}
and $\tau,\,\gamma \in \rs_+$. The corresponding transfer function is
\begin{equation}
    G(s) = C \left ( sI - A_0 - A_1 e^{-\tau s} - A_2 e^{-(\tau + \gamma) s} \right )^{-1}B.
    \label{eq:tdsfreq}
\end{equation}
The stability of this TDS varies depending on the values of the delays and has been analysed in \cite{niculescu2002delay}. Here, one considers $\tau = 1.2$ and $\gamma = 0.3$ for which \eqref{eq:tds} is stable.

\begin{figure}
    \centering
    \includegraphics[width=0.95\linewidth]{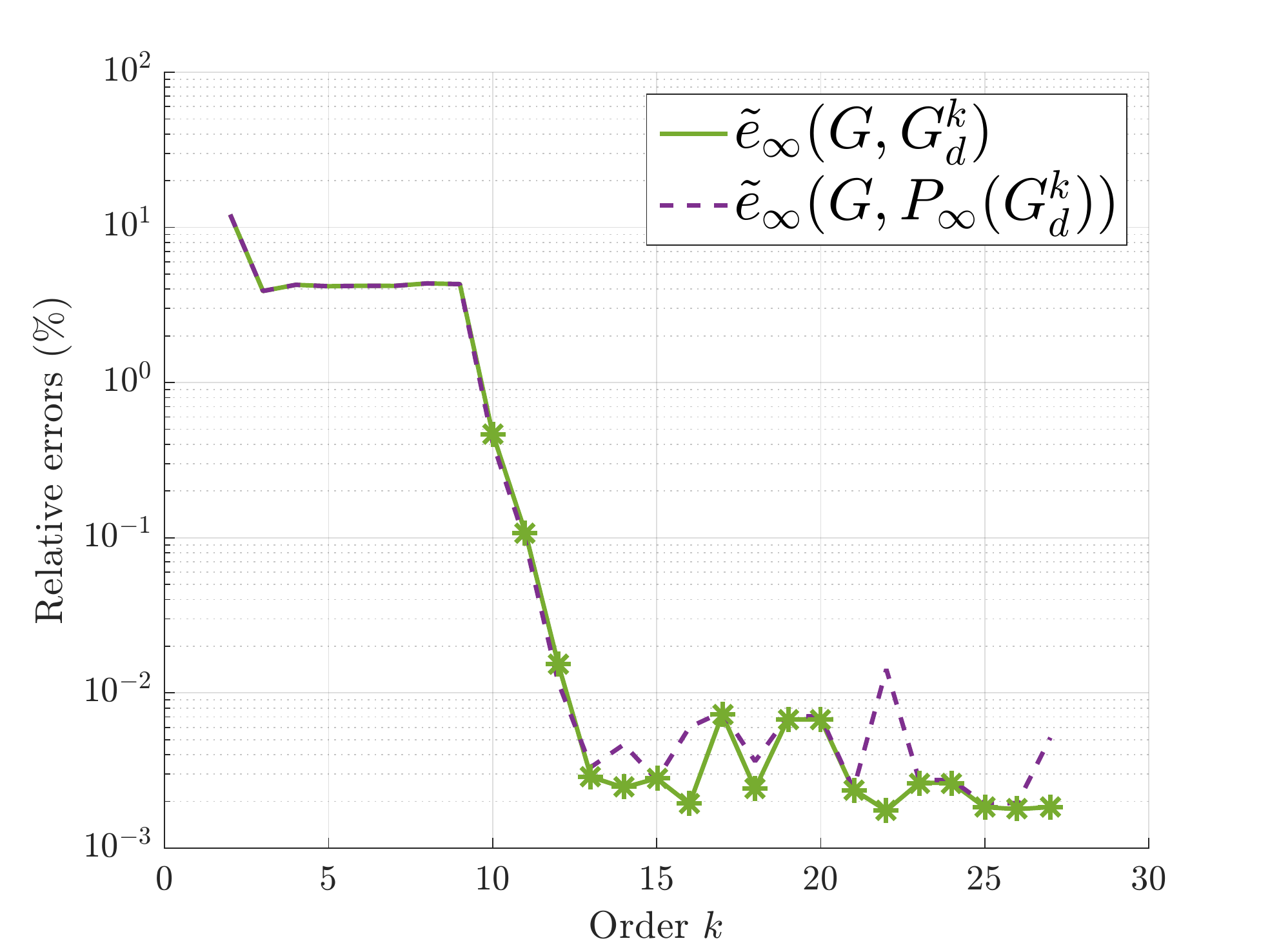}
    \caption{Relative frequency errors $\tilde{e}_\infty (G, G_d^k)$, $\tilde{e}_\infty(G, P_\infty(G_d^k))$ versus the order $k$ for the TDS case \eqref{eq:tdsfreq}.}
    \label{fig:tdsLoeErr}
\end{figure}

\begin{figure}
    \centering
    \includegraphics[width=0.95\linewidth]{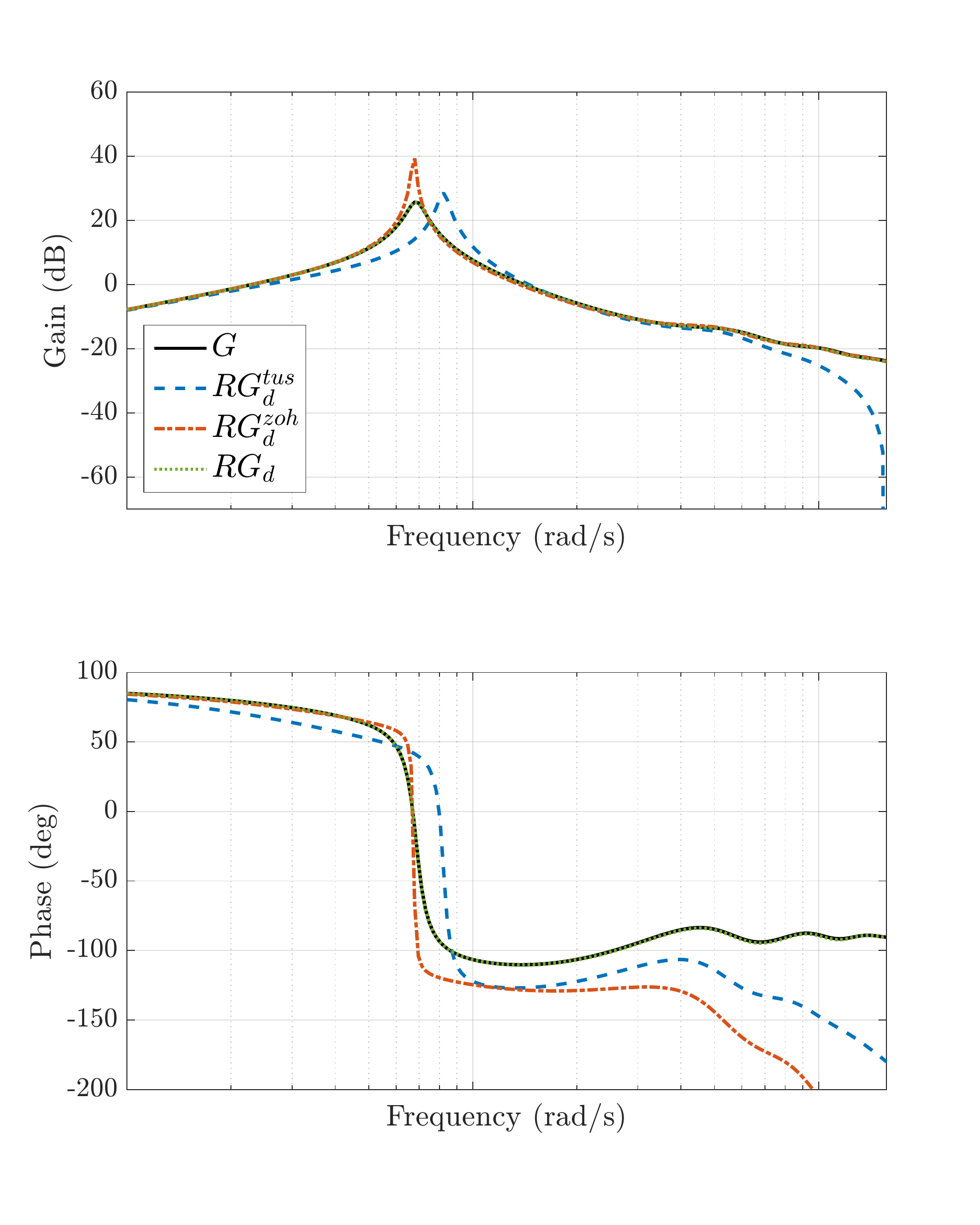}
    \caption{Frequency-domain responses of the time-delay system \eqref{eq:tdsfreq} and its discretised counterparts.}
    \label{fig:tdsBode}
\end{figure}

\begin{figure}
    \centering
    \includegraphics[width=0.95\linewidth]{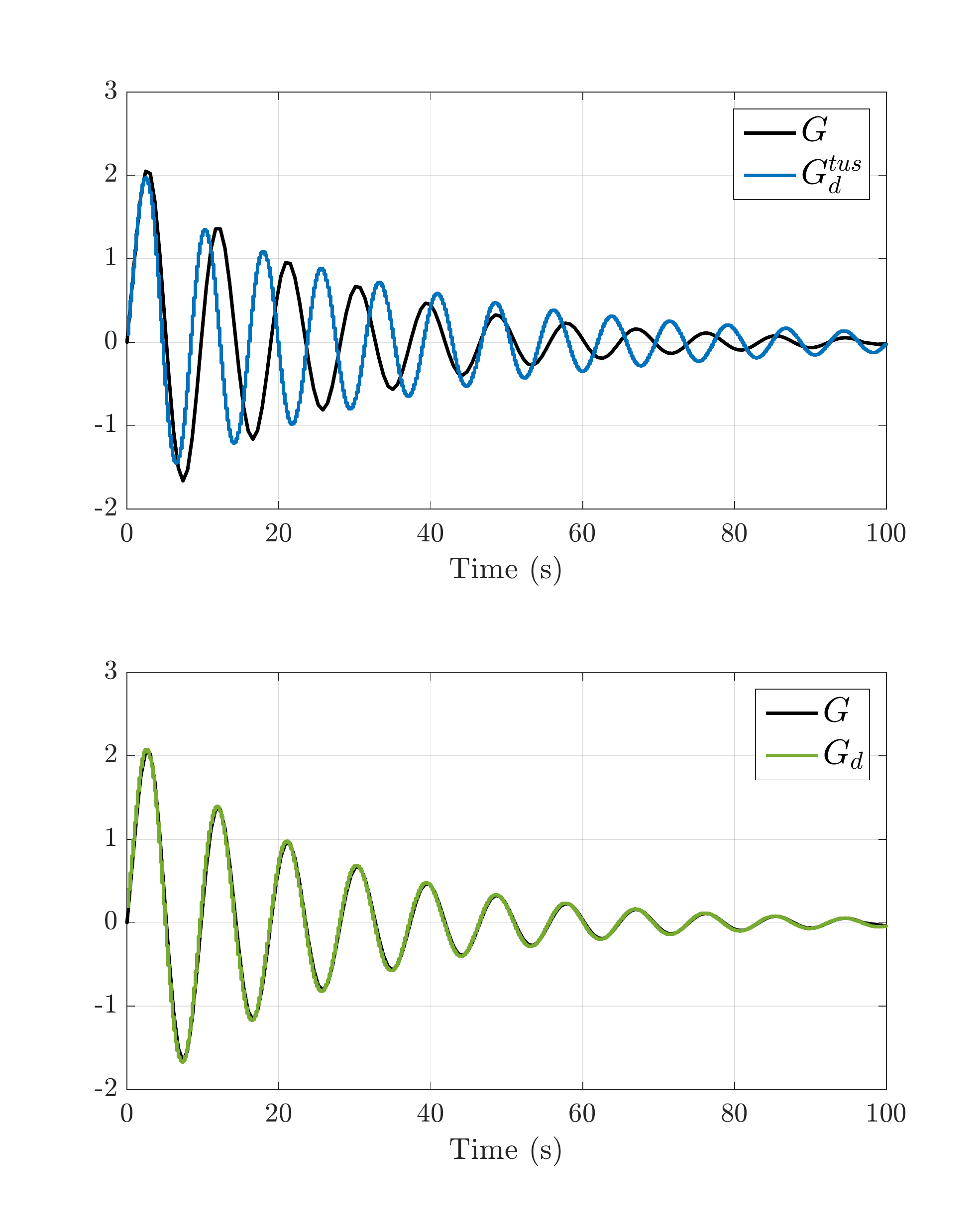}
    \caption{Step responses of the time-delay system \eqref{eq:tds} and its discretised counterparts $G_d^{tus}$ (top) and $G_d$ (bottom).}
    \label{fig:tdsStep}
\end{figure}

Considering the sampling period $h = 0.2$s, approximate discrete-time models are obtained with ZOH and Tustin methods\footnote{The impulse invariant method is not available for TDS.} with MATLAB. The resulting models $G_d^{zoh}$ and $G_d^{tus}$ are stable second-order models with internal delays. The Loewner approach is applied based on $100$ sample data linearly spaced in $\Omega = [10^{-3}, \omega_N - 10^{-3}]$ leading to a delay-free model $G_d^k$. As in the previous example, the relative errors obtained versus $k$ are reported in Figure \ref{fig:tdsLoeErr}, the frequency responses are plotted in Figure \ref{fig:tdsBode} where $G_d = P_\infty(G_d^{11})$ and the associated errors are reported in Table \ref{tab:TDS}. The step-responses of $G_d^{tus}$ and $G_d$ are also reported in Figure \ref{fig:tdsStep}. Note that $G_d$ is of order $10$, larger than $G$, but delay-free.

As in the previous example, the discretisation error decreases as the order $k$ increases but here instability appears later. Besides, the error between the interpolant model $G_d^k$ and its projection onto $\dhinf$ is negligible. In that case, increasing the order $k$ enables to embed the delayed nature of $G$. Both the frequency responses in Figure \ref{fig:tdsBode} and the errors in Table \ref{tab:TDS} highlight the efficiency of the proposed approach in comparison to standard methods. This is supported by the step response of Figure \ref{fig:tdsStep} where the response of the TDS model $G$ and of $G_d$ are superposed.

\begin{table}
\centering
\caption{Discretisation errors between the continuous-time TDS $G$ and its discretised counterparts.}
\label{tab:TDS}
\begin{tabular}{|c|cccc|}
  \hline
 $ model  $ & $G_d^{tus}$ & $G_d^{zoh}$ & $G_d^2$ & $G_d$ \\
 $\tilde{e}_{\infty}(G,\cdot) (\%)$&  $126.85$ & $503.71$ & $12.09$ & $\mathbf{0.094}$
 \\\hline
\end{tabular}
\end{table}



\section{Conclusion}
\label{sec:ccl}

An interpolation-based discretisation process for continuous-time LTI model has been proposed. It enables to modify the order of the discrete-time model to trade complexity for more accuracy without decreasing the sampling period. The potential of the approach has been highlighted on two numerical examples, including a Time-Delay System. Note that as the continuous and discrete time models play a mirror role in the approach, it is straightforward to adapt it so that it performs the converse and convert a discrete-time model into a continuous-time one.

For the TDS case, the resulting discrete-time model has no more delay. This can be an advantage as the discrete model belongs to the simpler class of rational models, albeit of potentially larger order, but it may be interesting for some applications to keep the delay explicit. This objective is under investigation and may be achieved by exploiting the delayed extension of the Loewner framework proposed in \cite{schulze:delay:2016}.

Current work also aims at evaluating the impact of this discretisation process for control and more specifically its impact for the performances of the closed-loop.





\bibliographystyle{plain}        
\bibliography{_biblioCPV,bibli}           

\begin{thebibliography}{10}

\bibitem{AntoulasBook:2005}
A~C. Antoulas.
\newblock {\em {Approximation of Large-Scale Dynamical Systems}}.
\newblock Advanced Design and Control, SIAM, Philadelphia, 2005.

\bibitem{aastrom:2013:computer}
K.J. {\AA}str{\"o}m and B.~Wittenmark.
\newblock {\em Computer-controlled systems: theory and design}.
\newblock Courier Corporation, 2013.

\bibitem{bamieh:lifting:1991}
B.~Bamieh, J.~Boyd~Pearson, B.A. Francis, and A.~Tannenbaum.
\newblock A lifting technique for linear periodic systems with applications to
  sampled-data control.
\newblock {\em Systems \& Control Letters}, 17(2):79 -- 88, 1991.

\bibitem{chen1991input}
T.~Chen and B.A. Francis.
\newblock Input-output stability of sampled-data systems.
\newblock {\em IEEE Transactions on Automatic Control}, 36(1):50--58, 1991.

\bibitem{chen1995optimal}
T.~Chen and B.A. Francis.
\newblock {\em Optimal sampled-data control systems}.
\newblock Springer Science \& Business Media, 1995.

\bibitem{Glover:1984}
K.~Glover.
\newblock {All Optimal Hankel Norm Approximation of Linear Multivariable
  Systems, and Their $\mathcal{L}_\infty$ error Bounds}.
\newblock {\em International Journal Control}, 39(6):1145--1193, 1984.

\bibitem{gosea:stability:2016}
I.V. Gosea and A.C. Antoulas.
\newblock Stability preserving post-processing methods applied in the loewner
  framework.
\newblock In {\em Workshop on Signal and Power Integrity}, pages 1--4, 2016.

\bibitem{izmailov:analysis:1996}
R.~Izmailov.
\newblock Analysis and optimization of feedback control algorithms for data
  transfers in high-speed networks.
\newblock {\em SIAM journal on control and optimization}, 34(5):1767--1780,
  1996.

\bibitem{Kohler:2014}
M.~Kohler.
\newblock {On the closest stable descriptor system in the respective spaces
  $\mathcal{RH}_2$ and $\mathcal{RH}_\infty$}.
\newblock {\em Linear Algebra and its Applications}, 443:34--49, 2014.

\bibitem{kowalczuk:discrete:1993}
Z.~{Kowalczuk}.
\newblock Discrete approximation of continuous-time systems: a survey.
\newblock {\em Circuits, Devices and Systems}, 140(4):264--278, 1993.

\bibitem{mari:modifications:2000}
J.~Mari.
\newblock Modifications of rational transfer matrices to achieve positive
  realness.
\newblock {\em Signal Processing}, 80(4):615--635, 2000.

\bibitem{Mayo:2007}
A~J. Mayo and A~C. Antoulas.
\newblock A framework for the solution of the generalized realization problem.
\newblock {\em Linear Algebra and its Applications}, 425(2):634--662, 2007.

\bibitem{niculescu2002delay}
S.~Niculescu.
\newblock On delay robustness analysis of a simple control algorithm in
  high-speed networks.
\newblock {\em Automatica}, 38(5):885--889, 2002.

\bibitem{schulze:delay:2016}
P.~Schulze and B.~Unger.
\newblock Data-driven interpolation of dynamical systems with delay.
\newblock {\em Systems \& Control Letters}, 97:125 -- 131, 2016.

\end{thebibliography}

\end{document}